# Optical Experiments on a Crystallizing Hard Sphere - Polymer Mixture at Coexistence

Andreas Stipp; Hans-Joachim Schöpe, Thomas Palberg,

Johannes Gutenberg Universität, Institut für Physik, Staudingerweg 7, D-55128 Mainz, Germany

Thomas Eckert and Eckhard Bartsch

Albert Ludwig Universität, Institut für Physikalische Chemie, Albertstraße 21, D-79104 Freiburg, Germany

Ralf Biehl

Institut für Festkörperphysik, Forschungszentrum Jülich GmbH, Jülich, Germany

**We report on the crystallization kinetics in an entropically attractive colloidal system using a combination of time resolved scattering methods and microscopy. Hard sphere particles are polystyrene microgels swollen in a good solvent (radius $a$=380nm, starting volume fraction 0.534) with the short ranged attractions induced by the presence of short polymer chains (radius of gyration $r_g$ = 3nm, starting volume fraction 0.0224). After crystallization, stacking faulted face centred cubic crystals coexist with about 5% of melt remaining in the grain boundaries. From the Bragg scattering signal we infer the amount of crystalline material, the average crystallite size and the number density of crystals as a function of time. This allows to discriminate an early stage of conversion, followed by an extended coarsening stage. The small angle scattering (SALS) appears only long after completed conversion and exhibits Furukawa scaling for all times. Additional microscopic experiments reveal that the grain boundaries have a reduced Bragg scattering power but possess an increased refractive index. Fits of the Furukawa function indicate that the dimensionality of the scatterers decreases from 2.25 at short times to 1.65 at late times and the characteristic length scale is slightly larger than the average crystallite size. Together this suggests the SALS signal is due scattering from a foam like grain boundary network as a whole.**

**Introduction**

Understanding the process that drives an undercooled fluid to solidification is still among the most challenging issues of condensed matter physics [1, 2]. It has been profitably approached by turning to colloidal suspensions, which can be regarded as effective one component systems where the disperse particles play the role of "macro-atoms". At constant temperature the nature, strength and range of interactions can be conveniently tuned [3]. Presence of the solvent on one side provides an effective heat bath and leads to isothermal phase transitions. On the other side it leads to diffusion governed dynamics restricting relevant time scales to the conveniently accessible, narrow range of a few milliseconds to several hours. Colloid specific length scales further allow for accessing system structure and dynamics as well as phase transition kinetics with optical techniques, like light scattering and microscopy [4, 5, 6, 7, 8]. For very concentrated samples also small angle x-ray or neutron scattering can be applied [9].

Concerning the choice of a suitable system for crystallization studies, the simplest experimental model consists of monodisperse hard sphere (HS) particles, e.g. polymer latices immersed in an organic solvent. For these crystallization is dominated by packing constraints, the volume fraction is the only control parameter and as for noble gases face centred cubic (fcc), hexagonally close packed (hcp) or randomly stacked hexagonal planes (r-hcp) result [10]. A coexistence region separates the fluid from the crystal state (with the volume fractions of freezing and melting at $\Phi_F = 0.495$ and $\Phi_M = 0.545$, respectively). At very large volume fractions a kinetic glass transition is encountered [5, 11].

Quantitative studies of solidification kinetics successfully employed both time resolved Bragg light scattering and in particular also Small Angle Light Scattering (SALS) [12, 13, 14, 15, 16, 17, 18, 19]. The Bragg signal yields information about the crystal structure, the lattice constant, the crystallite size and growth velocity, the fraction of crystalline material and the nucleation rate density [14]. The SALS signal for HS is due to the formation of a depletion zone about the crystallites showing supply limited growth. The crystals are compacted, while the depletion zone is diluted as compared to the remaining melt. As the number density of particles is conserved, a large scale fluctuation of the particle density and hence the refractive index results. This gives rise to a scattering signal at small scattering vectors $q$. The peak position corresponds to the average size of the combined object crystal plus depletion zone [20]. Thus this signal gives information about the growth mechanism and the extension of the depletion zone. In addition, it was observed that the peak shape reveals dynamical scaling, indicating a coupled evolution of crystallite size and depletion zone extension [12, 13, 14, 19]. Several scaling functions were proposed, which were derived from the classical Furukawa scaling form [21]. Their successful application showed that HS growth differs from late stage coarsening in spinodal composition, where also a peaked SALS intensity is observed [22]. Scaling was found to apply only up to the moment, when the depletion zones of neighbouring particles start overlapping. Then either scaling ceased or was interrupted and the scaling behaviour at later times followed still different scaling relations [16].

The addition of a non-adsorbing polymer to a HS system leads to an additional attractive term in the potential of mean force [23, 24, 25]. This term is of entropic origin and may be thought of as the local polymer osmotic pressure driving particles together. For sufficiently large polymer concentration a wealth of additional equilibrium and non-equilibrium phases is observed including stable colloidal liquids, a gel phase and an attractive glass [3, 26, 27, 28, 29]. Crystal formation has been frequently observed, and interesting studies about possible conversion scenarios have been reported [7, 30, 31, 32]. A quantitative study of the crystallization kinetics was, however, so far missing.

The richness of phase behaviour evolves gradually from the HS behaviour as the polymer concentration is increased. At low polymer volume fraction, the phase behaviour remains practically unchanged as compared to the reference HS system without polymer [33, 34]. In a recent letter we have, however, shown that the crystallization kinetics is already drastically altered as compared to the HS reference system [35]. A $t^{1/3}$ growth law was observed for both conversion and coarsening stage from the combined Bragg and SALS data. The SALS signal in addition showed dynamic scaling, following Furukawa´s law [21] after it appeared at late times, long after completed solidification. Such a kinetic behaviour is in agreement with theoretical expectation for a conserved order parameter [36, 37, 38, 39]. It occurs as compacted crystals are formed and the polymer is diffusively expelled to the surrounding fluid, to balance the osmotic pressure. The process continues also during coarsening and therefore the polymer density may take the role of a conserved order parameter. Inclusion of the polymer into the description of the crystallisation kinetics goes beyond the convenient effective one component model and points at a genuine many body effect.

As compared to the HS reference system, the different crystallization mechanism leads to significantly different scattering patterns at small angles, probing large scale fluctuations. In the present contribution we shall pay particular attention to the properties and the origin of this signal. In addition we give a more comprehensive account of the experiments reported in our letter, present an analysis of the Bragg scattering data obtained in parallel to the SALS data and give detailed account of the accompanying microscopic studies. In what follows we first introduce our sample and its conditioning, then shortly sketch the time resolved scattering experiments and the microscopic techniques employed. The discussion focuses on the contrast mechanism inducing the SALS signal, and on the information contained in it about the mesoscale structure of the (partially) solidified sample. We close with a comparison to other systems, highlighting differences and similarities to repulsive colloidal systems and describing the morphological analogy of our grain boundary network to a wet, non-drying foam.

.

## 2    Experimental

**Sample characterization and preparation**

The colloidal particles, polystyrene (PS) microgel spheres, were synthesized via surfactant-free emulsion radical polymerization in deionized water, using styrene as monomer, potassium peroxodisulphate as initiator and 1,3-di-isopropenyl-benzene as cross-linker, with a cross-link density of 1/10 [40]. The particles are dried and redispersed in the good solvent 2-ethyl-naphthalene (2EN), where they swell to their maximum size governed by the degree of cross-linking. In our case the swollen radius is $a = 380\pm4$ nm with a polydispersity $\sigma \approx 0.06$ (as determined from static and dynamic light scattering). The corresponding value in the collapsed state is $a_0 \approx 300$nm and the swelling ratio of $S = a^3/a_0^3 \approx 2.1\pm0.1$. Both the density $\rho$ and the refractive index $\nu$ of 2EN are close to the literature values for polystyrene ($\rho_{PS} = 1.05$ g/cm$^{-3}$, $\nu_{PS}$ =1.590, $\rho_{2EN} = 0.992$ g/cm$^{-3}$, $\nu_{2EN}$ =1.599 at T = 293 K). Swollen particles are even better matched, as here the volume weighted quantities apply: $\nu_{Part} = \sqrt{0.48\nu_{PS}^2 + 0.52\nu_{2EN}^2} = 1.595$.

In principle, a hard sphere like interaction results, but electrostatic and entropic contributions may occur as well. A small amount of dissociable surface groups is present in our system from the ionic polymerization initiator. Their degree of dissociation in the organic solvent is, however, very low and furthermore at no stage in the process other ionic species enter to adsorb and charge the particles. Therefore we expect only slight deviations from a hard repulsion. Unnoticed in earlier, preliminary experiments [41], however, the dry powder contains some amount of low molecular weight PS-chains from synthesis. The polymer component could be separated after swelling by repeated centrifugation and decantation. The separated free polymer solution was dried and weighed, then redispersed in THF and passed through a gel-permeation chromatography (GPC) to obtain its molecular weight distribution shown in Fig. 1. This analysis returned a free polymer content of 1.18% of the total polystyrene mass. Further $M_N = 6.42 \; 10^3$ g/mol and $M_W = 1.074 \; 10^4$ g/mol, which corresponds to a radius of gyration of $r_g \approx 3$ nm in a good solvent. Purified dispersions of microgels in 2EN on the other side show typical HS phase behaviour, dynamics and rheology [33, 34].

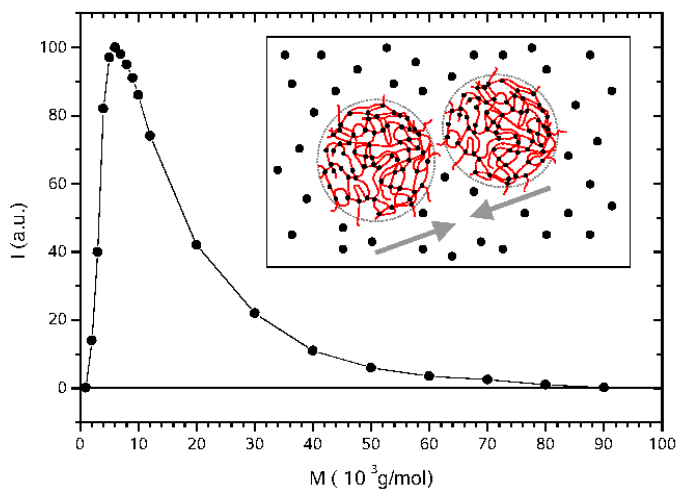

*Fig. 1: GPC molecular weight distribution of the free polymer chains in our sample. When two microgel particles are sufficiently close to each other to exclude the polymer coils from the region between them, they experience an effective depletion attraction (inset).*

Experiments reported in our previous Letter and in this study were performed using the particle batch *without* purifying. A drawn to scale sketch of the dispersion components is given in the insert of Fig. 1 for a number ratio of 1:20. Extrapolating the trends observed in previous experimental systems of larger size ratio to our value of $\chi = a_{POLY}/a = 0.008$ ($a$ = radius of microgel particles, $a_{POLY} = r_g$ of the polymer), we expect such a combination to yield a strong but rather short ranged attraction due to the depletion effect. The nominal volume fraction was chosen to be $\Phi = 0.54$, i.e. at the upper end of the fluid-crystal coexistence region. The true volume fraction is lower, if we account for the 1.18% w/w dissolved polymer. The mass of an individual particle is $m_{Part} = \rho_{PS} (4\pi/3) a_0^3 = 1.188 \; 10^{-16}$ kg, the mass of a polymer is $m_{Poly} = M_n/N_A = 1.07 \; 10^{-23}$ kg. The resulting number ratio then is $n_{Part} / n_{Poly} = (0.9882 / m_{Part}) / (0.0118 / m_{Poly}) = 7.5 \; 10^{-6}$, where $n$ denotes the number density. The volume fraction ratio is $\Phi_{Part} / \Phi_{Poly} = n_{Part} V_{Part} / n_{Poly} V_{Poly} = 15.3$. Therefore a nominal volume fraction of 0.54 corresponds to an experimental volume fraction of $\Phi_{Part} = (1-0.0118) \; 0.54 = 0.534$ and a polymer volume fraction of $\Phi_{Part} / 15.3 = 0.035$. The corresponding volume fraction based on the weight fraction is 0.0224 [35].

The sample was mixed from weighed quantities of dispersion powder and solvent and filled into the sample cell. It was left undisturbed for several weeks, waiting for the equilibration of the microgel swelling. It finally appeared full of small crystals, with no detectable sedimentation effects. The crystals could be molten again by applying a moderate shear stress i.e. by gently tumbling the sample, but readily re-crystallized within some 10min.

**Scattering experiments**

The crystallization process was monitored *via* light scattering experiments. We employed a machine originally designed by K. Schätzel to simultaneously measure the Bragg and the small angle light scattering of PMMA hard spheres. The interested reader is encouraged to see the detailed description in [14]. In Fig. 2 we show a schematic drawing of our set-up

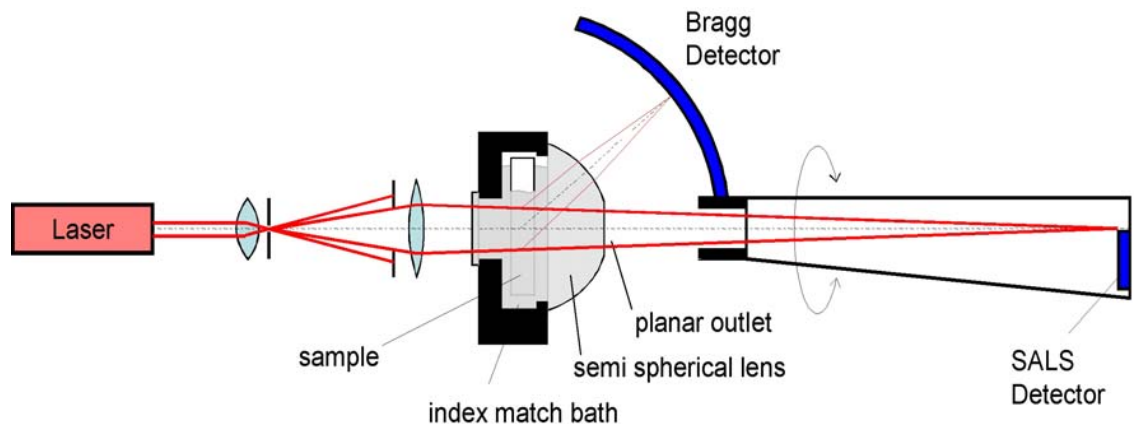

*Fig. 2: Schematic drawing of the combined Bragg- and Small angle scattering experiment.*

The sample is contained in a cell of rectangular cross section of $(5\times10)mm^2$ and placed in an index match bath containing 2EN to avoid parasitic reflections. All glass-ware (of cell, match bath and detection optics were tailor made to have a refractive index very similar to that of 2EN in order to avoid any refraction of light except at the front and rear windows. This is shown in the drawing by using the same light gray colouring except for the upper air-filled part of the cuvette. The sample is illuminated by a widened He-Ne-laser beam (∅ 8mm) focused onto the centre of the small angle detector some 1.5m behind the sample cell. The laser beam centre defines the optical axis. The plane front (inlet) window of the match bath is oriented perpendicular to the optical axis, while the rear (outlet) window is a hemispherical lens with a plane part at its centre. The lens focuses the 2D scattering pattern of Bragg scattered light I(Θ,φ) onto points on a spherical focal plane. There it is collected by the Bragg detector comprising of arrays of photo diodes mounted on a curved, rotating arm. During measurements the sample is left at a fixed position, while the detector is rotated about the optical axis. This averages over the 2D scattering pattern and yields I(Θ). Our approach has the advantage of minimizing mechanical impact on the fragile colloidal crystals, while keeping good statistics. A full rotation takes some 30s and defines the maximum temporal resolution of the instrument. We chose a time step of 1min for the experiments at early times. Before the measurements the sample was kept in a shear molten state on a tumbling device. A time lag of about 1min separates the cessation of shear (with the sample being on the tumbler) and the completion of the reference measurement (with the sample aligned in the machine). Start of the following measurement defines t=0.

Also for SALS the detector is placed on a rotating arm, now using the inner, plane part of the rear window for the outlet of scattered light and laser beam. The range of accessible scattering vectors $q = (4\pi v/\lambda)\sin(\Theta/2)$ (where $\lambda = 633$nm is the laser wave length in vacuum and Θ is the scattering angle) is considerably different for the two experiments. For Bragg light scattering $5.2\mu m^{-1} \leq q \leq 19.3\mu m^{-1}$ covering the first five reflections and for SALS $0.028\mu m^{-1} \leq q \leq 0.41\mu m^{-1}$ covering the majority portion of the small angle peak with a resolution of $0.01\mu m^{-1}$.

In Fig. 3a we show the temporal evolution of the Bragg peaks in terms of the scattered intensity divided by the scattered intensity measured in the reference measurement ending at t = 0. Five peaks corresponding to the (expected) face centred cubic structure of the crystals are clearly identifiable. {200} and {222} are too small for quantitative evaluation. Both {111} and {311} show a sixfold scattering pattern in addition to the Debye Scherrer ring in the complete 2D scattering signal. Such a superposition has been observed also in other systems and is ascribed to the presence of a wall nucleated crystal [42]. {111} in addition shows a quite asymmetric shape also in the non-normalized data due to the formation of stacking faults [10]. This leaves the {220} reflection for evaluation, which displays a comparably smooth background and a symmetric peak shape. Apart from constant experimental parameters, the scattered intensity depends on the density of scatterers $n$, their scattering strength $b_0^2$, the particle form factor $P(q)$ and the structure factor $S(q,t)$ which evolves from fluid to crystalline during solidification: $I(q,t) \propto n\, b_0^2\, P(q)\, S(q,t)$. We divide by the earliest measurement to eliminate $n\, b_0^2\, P(q)$ and subtract a linear background function[1] to isolate the scattering contribution of the crystalline phase in the vicinity of the {220} reflection. Then a Lorentzian is fitted to the resulting peak from which the position of the maximum $q_{MAX}(t)$, the full width at half height $\Delta q(t)$ and by integration the area under the peak $A_{220}(t)$ are inferred. We note that the choice of this fitting function is not motivated by physical reasoning, but rather facilitates a convenient parameterization. Also, {220} is a relatively small peak. This limited the duration of the Bragg experiment in the sense that after some 2200min the peak width became too small for a reliable fit of the Lorentzian, although data were taken up to 8500min.

The small angle raw data are shown in Fig. 3b. For t = 1min a monotonously decreasing intensity is observed. For the smallest q-values this signal (red curve) decreases over the first 15 to 30 min to stay practically unchanged for the next two hours (blue curve). We therefore use the signal at 93min as reference background for our suspension after the end of its conversion stage. If this background is subtracted from the following spectra (c.f. Fig. 3c) one observes a small angle signal to evolve out of the noise after some 200min. It shows a single broad peak which can be followed with sufficient resolution up to some 8000min. With increasing time the peak gains intensity and the position of the maximum shifts towards smaller q.

---

[1] Lacking a more advanced knowledge of the actual course and development of the scattering contribution of the amorphous phase, we here for convenience approximate this quantity by a linear function

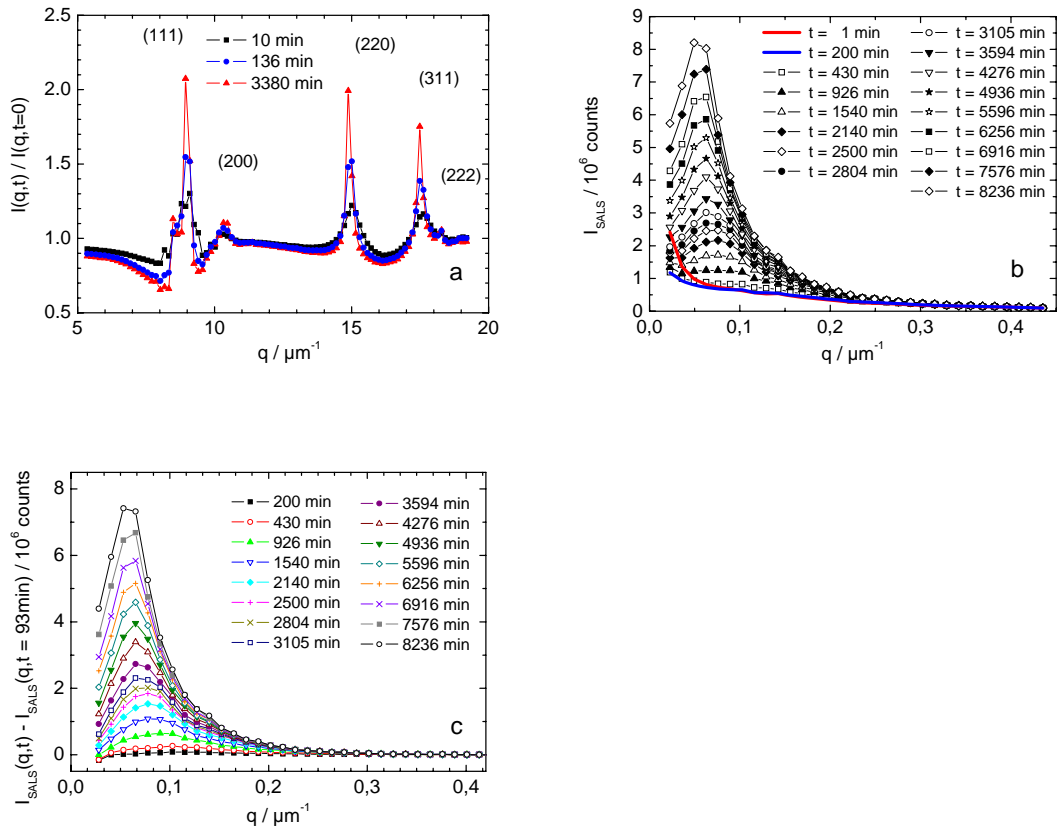

*Fig. 3 scattered intensities: a) evolution of the five Bragg reflections (Miller indices as indicated) as a function of time; b) raw small angle data as a function of time c) evolution of the small angle signal after back ground (=measurement at 93 min) subtraction as a function of time .*

## Microscopy

In addition to the scattering experiments we performed microscopic experiments to study the sample morphology and to investigate the origin of the small angle signal. To be specific, the sample morphology was studied using Bragg microscopy, in which the sample is illuminated with white light from aside. The Bragg scattered light was collected by a CCD camera equipped with a macro lens. Crystallites appear as colourful polygons with the colour depending on their orientation.

The origin of the small angle signal was investigated by both phase contrast microscopy (PCM) and differential interference contrast microscopy (DICM). In both cases the suspension was filled into a rectangular quartz cell of $2.5 \times 10 mm^2$ cross section and (after tumbling) placed wide side up on the stage of an inverted microscope (IRB Leica, Germany), where it solidified. We used 20fold magnification objectives with numerical apertures of 0.4. Images showing good contrast could be recorded after some 500min using a digital camera with 5Mpix. resolution. In our case particles, polymer and solvent are non-adsorbing in the visible. Still, in transmission the colloidal solid may act as an effective amplitude object, since both crystalline and fluid regions Bragg scatter at larger angles, which are not captured by the objective and thus the transmitted light is attenuated. In addition, due to the differences of composition in grain boundaries and in crystals, these have a different refractive index, and thus are phase objects, too. Details of both techniques with application to the case of colloidal dispersions have been given in the excellent review of Elliott and Poon [43]. A brief introduction into both techniques is given in the appendix. We here only note, that the combined use of these two techniques allows to characterize the differences in refractive index and scattering power between crystallites on one and grain boundaries and melt pockets on the other side.

## 3      Results

**Time resolved Bragg scattering**

We show the evaluation of the time dependent scattered intensities in the Bragg and in the SALS regime in Figs. 4a-d. A convenient measure for the time scale of the observed phase transition kinetics is given by the Brownian time $t_B = a^2 / D_0$, where $a$ is the particle radius and $D_0 = 2.22$ $10^{-13}$ m$^2$s$^{-1}$ is the free diffusion coefficient in 2EN. Then $\tau = t / t_B$, with $t$ measured in s, is a dimensionless time and 1 min correspond to $102.7\tau$. Both time scales are shown in Fig. 4a as upper and lower x-scales.

Fig. 4a shows the integrated Bragg intensity $A_{220}$ which initially increases with an increasing slope. Its increase slows after about seven minutes and from about 10min $A_{220}$ stays roughly constant for the rest of time. From t = 2min the full width at half height $\Delta q$ decreases continuously and with decreasing slope. The upper curve in Fig. 4a shows the evolution of the 220 peak position. It decreases from $q_{220}(t=0) = 15.2 \mu m^{-1}$ to $q_{220} = 14.87 \mu m^{-1}$ at very long times, with a period of faster decrease between 5 and 10min.

From these data we now may extract the physical parameters of the crystallization process. First we calculate the fraction of crystalline material $X_X(t)$, obtained as $A_{220}(t)$ normalized by its long time value and multiplied by 0.95, the final crystallinity estimated from the microscopic analysis (see section "Microscopy" below). A double logarithmic plot of $X_X(\tau)$ versus $\tau$ yields a power law behaviour $X_X \propto \tau^\alpha$ with an exponent of $\alpha_X = 1.33 \pm 0.03 \approx 4/3$ for the first seven minutes. The data are shown in Fig. 4b together with a straight line of slope 1.33 as a guide to the eye. $X_X(t)$ saturated for times larger than some 10 minutes. Following earlier literature [12, 13, 14, 15], we therefore discriminate an early stage of conversion and a later stage of coarsening and mark this transition by the vertical dotted line in Fig. 4b.

Next we calculated the characteristic dimensions of scattering objects. Here the average side length of a crystallite modelled as a cube $<L>$ is used, given by $<L> = 2\pi K / \Delta q$, where the Scherrer constant $K$ arises from geometrical considerations and is of order unity [44]. A double logarithmic plot of the data versus $\tau$ yields a power law behaviour $<L> \propto \tau^\alpha$ with an exponent of $\alpha_L = 0.28 \pm 0.015 \approx 1/3$ for all times up to $5 \cdot 10^4 \tau$. A slight decrease of the exponent is seen at larger times. Interestingly, $<L>$ does not increase significantly during the second measurement, while at the same time the crystallinity increases by a factor of three. This resembles the two step crystallization recently observed in experiments on polydisperse HS [45], but also found in NPT Molecular Dynamics simulations on monodisperse HS [46]. In both cases, first densified clusters are formed, which later develop fully crystalline order (hcp, fcc and/or r-hcp). Our temporal resolution is, however, insufficient to draw more definite conclusions.

During crystallization, the peak position shifts to smaller values. Using $g = 2\pi/q \, (h^2+k^2+l^2)^{1/2}$, where $h$, $k$ and $l$ are the Miller indices, to calculate the lattice constant $g$, and further $\Phi = 16\pi a^3/3g^3$ with $a = 380$nm we obtain a starting volume fraction of $\Phi = 0.557$ and a final volume fraction of $\Phi = 0.540$. Within the experimental errors of the particle radius and the peak position, the latter value is close to the theoretical value of the pure hard sphere melting volume fraction $\Phi_M = 0.545$.

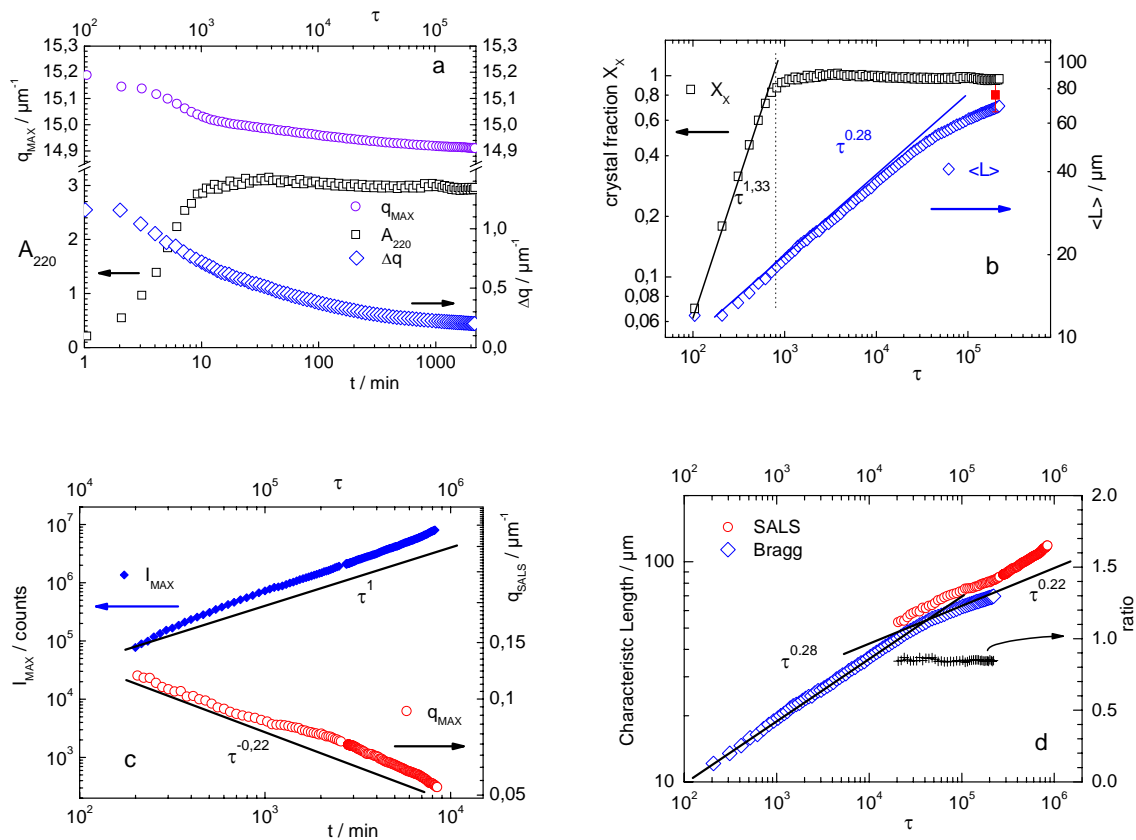

*Fig. 4: Evaluation of Bragg and SALS data. a) Bragg data. Open circles: position of the Bragg peak $q_{MAX}$ (upper left scale); open squares: integrated scattering intensity $A_{220}$ (lower left scale); open diamonds: full width at half height $\Delta q$ of the Lorentzians fitted to the Bragg peak (right scale). Data are shown up to 2200min corresponding to $2.3\ 10^5\ \tau$ (upper x-scale; $\tau = t/t_b$ with $t_b = 0.59s$, see text). For longer measurement duration the peak becomes too narrow to extract reliable data. b) Double log plot of the crystallite fraction $X_X$ (open squares) and the average crystallite size $<L>$ (open diamonds). For comparison we also show the average crystallite size determined from microscopy after 2000min (filled square). Lines of slopes 1.33 and 0.28 are guides to the eyes corresponding to the results of least square fits of power laws to the scattering data. Note that crystal sizes increase unaffectedly over the whole time range. c) SALS data. Filled diamonds: maximum intensity of the SALS peak $I_{MAX}$; open circles: position of the SALS maximum $q_{SALS}$. Lines of slopes 1.0 and -0.22 are guides to the eyes corresponding to the results of least square fits of power laws to the data. d) Comparison of characteristic length scales and power laws from Bragg scattering (open diamonds) and SALS (open circles). Crosses show the ratio of these values to be constant over the time range of overlap.*

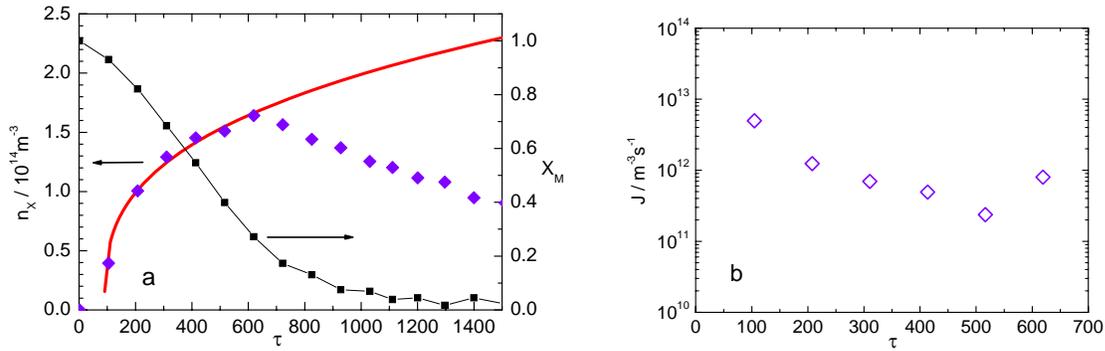

*Fig. 5: Evaluation for nucleation rate densities: a) closed squares: melt fraction $X_M$; closed diamonds: crystallite density $n_X$ as derived from the crystallite sizes $<L>$ and the crystal fraction $X_X$. The solid red line is a least square fit of a power law to the data returning an exponent of $\alpha_n$ = 0,35±0,05 and a "induction" time of $\tau_0 = (90±10)\tau \approx 1 min.$; b) nucleation rate density J as calculated for the initial stage of solidification assuming an induction time of $\tau_0 = 90\tau$. For $200\tau < t < 700\tau$ we find $J \approx 10^{12}$ nucleation events per second and cubic meter of remaining melt volume.*

Another interesting quantity that can be estimated from the Bragg data is the nucleation rate density $J$, giving the number of nucleation events per time and volume. This is illustrated in Fig. 5a,b. The fractional volume of the remaining melt is $X_M = 1 - X_X$ (filled black squares in a). The average volume of crystallites is given by $<L>^3$. From this and the crystallite fraction the crystallite number density $n_X = 10^{18} X_X / <L>^3$ (in m$^{-3}$) is calculated (filled diamonds in a) which first increases, then after seven minutes, starts to decrease. Crystallites may appear *via* continued homogeneous nucleation in the bulk melt. Crystallites may also disappear due to ripening processes. We therefore discriminate an early regime of dominant crystal nucleation and a later regime of dominant crystal ripening. Note however, that the cross-over time is smaller than the time of saturation in $A_{220}$ resp. in $X_X$. Thus in contrast to most earlier observations on pure HS ripening is already present in the late conversion stage. (For an interesting exception observed under µ-gravity conditions, which is possibly related with the two step nucleation process seen in polydisperse hard spheres [45], see [47].) In the nucleation dominated regime we can describe the data by a power law increase for $n_X$ with $\alpha_n = 0.35±0.05$ if we allow for a finite "induction" time of $\tau_0 = (90±10)\tau$ counted from the start of the first measurement. In the ripening dominated regime the data obey a power law with $\alpha_{ripen}$ = -0.85±0.05 up to $10^5\tau$. For larger times a smaller slope is observed also for $n_X$ parallel to the decreased slope of $<L(t)>$ (data not shown).

Crystallites can nucleate in the remaining melt only. To estimate a nucleation rate density we divide the difference in crystallite densities between two measurements by the remaining melt fraction $X_M$ and by the elapsed time in sec. For the first measurement, which ended 102.7$\tau$ = 60s after the reference measurement, the elapsed time was taken to be $t_1$ = 12,7$\tau$ ≈ 7.4s. All further time intervals are of 102.7$\tau$ = 60s. The thus estimated nucleation rate density is plotted in Fig. 5b. It is roughly constant on the order of $10^{12}m^{-3}s^{-1}$ perhaps with a slight tendency to decrease. We note, however, that the first point has an uncertainty of at least one order of magnitude due to the uncertainty in the induction time leading to a strong uncertainty in the elapsed time. Lacking sufficient time resolution, we may therefore not comment on these earliest stages of conversion, and e.g exclude or confirm an initial burst of nucleation as suggested in several recent studies on pure HS [45]. The other, later data points have uncertainties of about a factor of two. Thus we may conclude that no strong variation of the nucleation rate density with time is observed between two and six minutes (200-1200$\tau$). I.e. over most of the conversion regime steady state nucleation prevails.

A consistency check on the crystallization scenario at early times can be performed considering that the total amount of crystalline material is given by the number of crystallites times their average volume. In fact, $\alpha_X$ = 1.33 while $\alpha_n + 3\alpha_L$ = 1.19. Both values agree within roughly 10% and the difference is covered by the residual uncertainties of the fitted exponents. We thus can conclude that the microgel plus polymer system nucleates at an approximately constant rate density of $J \approx 10^{12}s^{-1}m^{-3}$, while at the same time the crystallite sizes increase with the cube root of time. At later times this increase is retained with no detectable change across the transition from the nucleation to the ripening regime nor across the transition from the conversion to the coarsening stage.

**Time resolved small angle scattering**

SALS intensity is present already in the first measurement. The signal shows a monotonically decreasing intensity and relaxes towards a stationary signal (within experimental uncertainty) over a time interval of 15 to 30 min. This is shown in detail in a semi-logarithmic plot in Fig. 6a.

Apart from the continuous decrease of intensity in the first two channels (which appears to continue for the first two hours) the decrease is most pronounced in for the q-range of $0.1\mu m^{-1}$ to $0.2\mu m^{-1}$. The observable changes are, however, tiny and way below the effects seen in pure HS systems, where already in the earliest stages of conversion a pronounced peak evolves due to depletion zone scattering. Here the feature disappears.

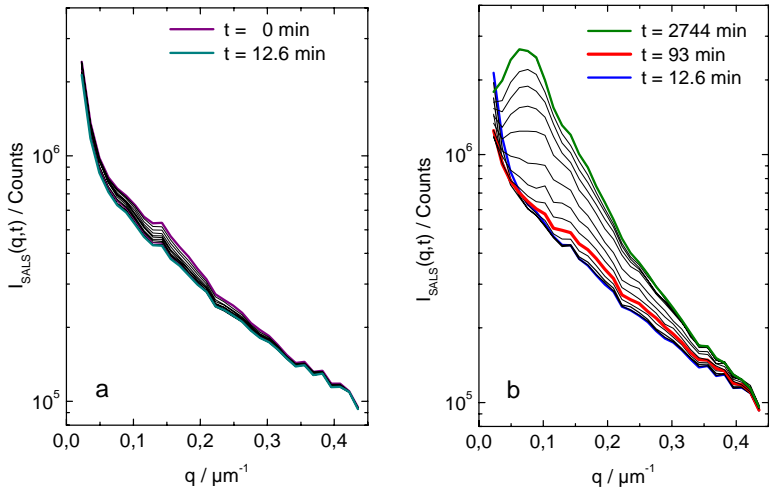

*Fig. 6: SALS intensities for different times: a) evolution from t = 1min to t = 12.6min (thick solid lines) in time intervals of approximately 1min; b) evolution from 12.6min to 2744min for increasing time intervals. The red solid curve shows the signal chosen as reference for the system after completed conversion.*

Rather, we observe the SALS peak to become a significant deviation from the stationary signal only long after the end of the conversion stage at $t = 7$min. In Fig. 6b one observes that between 12.6min and 93min only a tiny intensity increase occurred in the $q$-range of $0.1\mu m^{-1}$ to $0.3\mu m^{-1}$, which is still only on the order of the experimental resolution. At $t = 2744$min a pronounced peak has evolved. To isolate this peak signal, we subtracted the measurement at $t = 93$min (just before the peak-like signal exceeds the noise background) serving as reference for the fully converted state from all later measurements (c.f. Fig. 3c).

The isolated peak was subjected to further analysis. In the double logarithmic plot of Fig. 4c we show that the peak intensity $I_{MAX}$ increases with a power law in time: $I_{MAX} \propto \tau^\alpha$ with $\alpha_{I,SALS} \approx 1$. At early and at late times the increase is slightly larger. The position of the small angle maximum is shifting towards lower $q$. Again a power law behaviour is observed with $\alpha_{q,SALS} = 0.22\pm0.02$ for the first decade in time. Also for SALS a characteristic length can be determined from the position of the maximum as $L_{SALS} = 2\pi / q_{MAX}$. Fig. 4d compares the results for the characteristic lengths of SALS and Bragg LS with each other. A close agreement of the power laws for the increase of the characteristic size is observed. In particular for times between $2 \, 10^4 \tau$ and $2.2 \, 10^6 \tau$ the ratio of the two values stays remarkably constant at a value of 0.85±0.05. Hence the same curve shape is observed for both curves. This strongly suggests that the SALS signal is related to the crystallite size.

At 2000min the crystallite size from Bragg scattering appears to be somewhat smaller than the size obtained from microscopy ($<L>_M$ = (76±9)µm). Both are, however, smaller than the characteristic length inferred from the SALS signal for all overlap-times. The first observation may be due to a broadening of the Bragg peak caused by a slight misalignment of the Bragg detectors with respect to the focal sphere created by the lens. Secondly, an additional broadening may be expected from crystal straining occurring upon intersection and due to particle rearrangements during coarsening. For each individual crystal this will shift the peak position according to the lattice distortion. As a result of averaging over many crystals $\Delta q$ may increase leading to an apparently smaller crystallite size. The second observation, that the SALS signal yields a larger characteristic length than the Bragg signal and the microscopy results, indicates that it arises from slightly larger objects, which are, however, related to the crystallite size.

Generally, a peak in the small angle regime corresponds to large scale fluctuations of the refractive index which are typically caused by density fluctuations. This has formerly been observed for crystallizing hard spheres suspensions [12, 13, 14, 16] but also in spinodal decomposition of (atomic or molecular) binary liquid mixtures [36]. Such peaks were observed to show a scaling behaviour, i.e. they collapse on a master curve, if the intensity $I(q)$ is normalized by the maximum intensity $I_{MAX}$ and the scattering vectors $q$ by the position of the maximum $q_{SALS}$. In condensation from the vapour phase and in many binary mixtures the functional form could in addition be described by Furukawa´s function $F(q) = I(q)/I_{MAX} = (1+\delta/2)Q^2 / (\delta/2 + Q^{\delta/2})$, where $Q = q/q_{SALS}$ and $\delta$ is related to the fractal dimension of the scattering objects $d_f$ as $\delta = d_f + 1$ [21]. This function relates the distances of the scattering objects with their shape and size. Scaling then indicates a co-evolution of both quantities and thus a self similar morphology. In hard spheres systems this form was not met or only over rather restricted ranges of time, and other scaling functions have been proposed [12, 13, 14, 16].

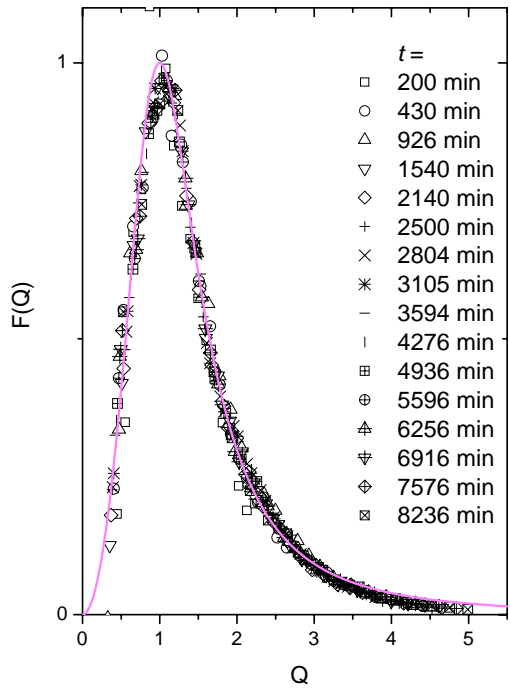

*Fig. 7: Small angle data plotted in terms of the scaled intensity $F(q) = I(q)/I_{MAX}$ versus the reduced scattering vector $Q = q/q_{SALS}$. All data collapse on a single master curve, which is well described by a Furukawa scaling function with $\delta = d_f + 1 = 3$ (solid line). Small deviations are seen for data taken at very early times and for all data at large Q, where the signal is only weak.*

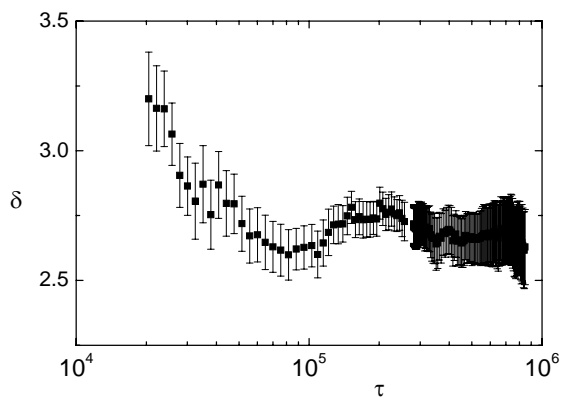

*Fig. 8: Semi-log plot of the Furukawa fit parameter $\delta = d_f + 1$ versus time. After an initial fast decrease $\delta$ levels of at a value of approximately 2.65.*

In Fig. 7 we show, however, that our SALS data all collapse on a single master curve if plotted in terms of $F(q)$ versus $q$. Moreover, the master curve is well described with Furukawa´s function over the complete range of reduced scattering vectors. $F(Q)$ is shown as solid line for comparison, with $\delta = d_f +1 = 3$. Deviations are mainly visible at very large $Q$ where the experimental SALS signal was already very weak. Also the very first data set deviates. We therefore fitted each individual data set with Furukawa´s formula taking $\delta$ as a free parameter. In Fig. 8 we show that $\delta$ drops quickly from values about 3.25 for the first measurements to level off at an average value of 2.65 for times larger than $10^3$min or $10^5\tau$. Thus the Furukawa scaling fit yields a fractal dimensionality of 2.25 decreasing to 1.65 for the small angle scattering objects. For a further identification of the scattering objects we performed a microscopic analysis.

**Microscopy**

Fig. 9 shows a Bragg microscopic image of the sample during an early stage of ripening ($t \approx$ 50min). The scale bar is 100µm. The sample was contained in a cylindrical vial, which causes the blurring at the sides of the image recorded using the digital camera with a macro lens. One observes a mosaic of irregular polygons of up to 30µm in diameter. Their corners appear rounded and the sides are often somewhat irregular and curved. Some of the crystallites show zebra stripes corresponding to stacking faults causing a twinning between ABCABC and ACBACB stacking of fcc.

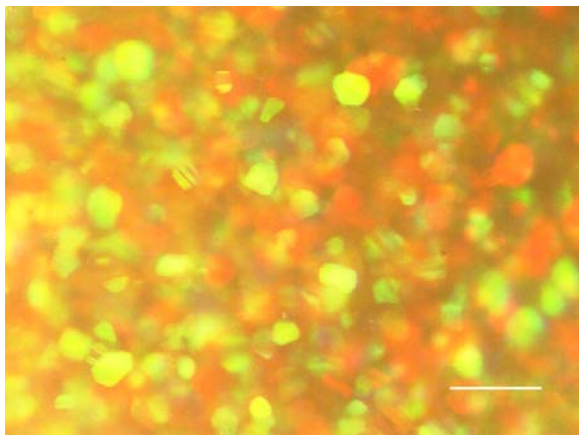

*Fig. 9: Bragg micrograph of the sample taken after 50min. The scale bar is 100µm.*

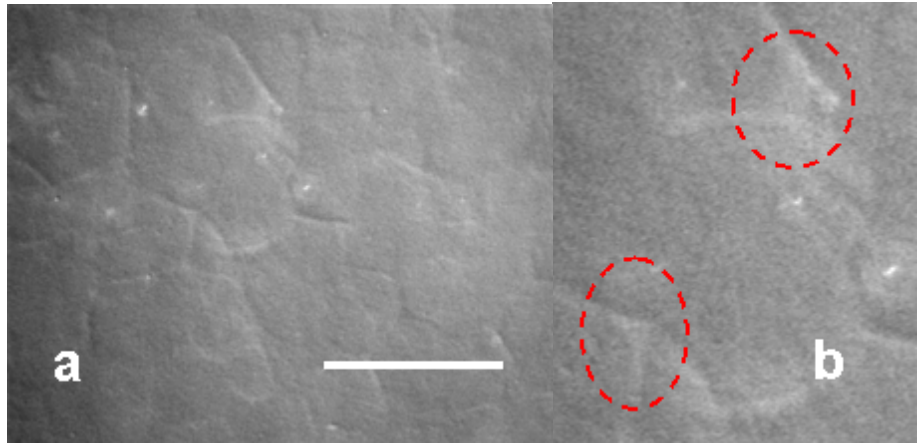

*Fig. 10a,b: DICM micrograph of the sample after 2000min. Note the characteristic relief-like appearance. Left: full image. The scale bar is 100µm. Right: close up with two pockets of remaining melt marked by dashed lines.*

Fig. 10a shows a DICM image of the sample recorded after 2000min. The scale bar is 100µm. A line pattern is visible. The lines correspond to the positions of the grain boundaries and appear white at the left and lower side, while the upper and right sides appear dark, resulting in a relief like image and corresponding to amplitude gradients of opposite sign at either side of the grain boundaries. An analysis of the intensity along a direction normal to the lines shows that on average the deviation of the bright part from the average value is considerably larger than the negative deviation of the dark side. This shows that in addition phase gradients are present on either side of the grain boundaries. DICM, however cannot discriminate the sign of the phase difference. Two particularly interesting parts of the picture are enlarged in Fig. 10b and marked by dashed lines. Here 10 to 30 micron wide, slightly lighter homogeneously grey areas are visible which are positioned at the intersections of grain boundaries. They correspond to amplitude objects with somewhat less attenuation than their surroundings and can be identified as pockets of remaining melt. As our system is non-adsorbing we observe effective amplitude objects, where the extinction is caused by scattering. We thus conclude that at $t = 2000$min the grain boundaries and melt pockets Bragg-scatter less than the crystallites.

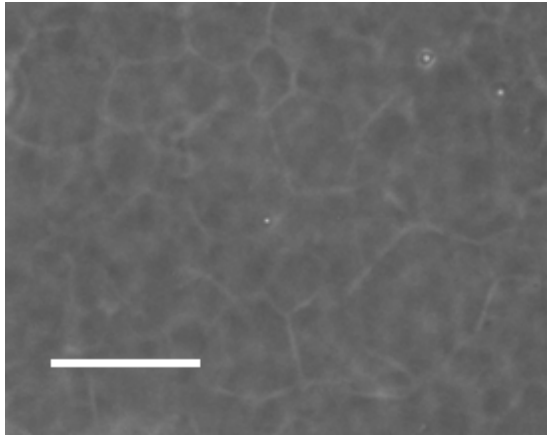

*Fig. 11: PCM micrograph of the sample taken after some 2000min. The scale bar is 100µm. Fine light grey lines appear on a darker grey background.*

Fig. 11 shows the corresponding PCM image. Here a network of light grey lines is seen on a darker grey background. The crystalline regions in addition show a mild, rather smooth variation of darkness. An increased intensity in PCM with positive phase contrast is compatible both with a increased refractive index of that region as compared with the average refractive index and with a decreased attenuation as also seen in DICM. Line scans of the intensity normal to the grain boundaries however show that the contrast is considerably larger than that between the crystallites and the melt pockets seen in DICM. We therefore conclude that in addition to a decreased attenuation the grain boundaries possess an increased refractive index as compared to the crystallites, which both increases the grain boundary brightness.

In both Fig. 10 and Fig. 11 the lines appear to be straightened as compared to the early time image in Fig. 9. Also in several cases they show angles close to either 120° or 108° which are rare at early stages. In one case a right angle is seen. The average full width at half height of eleven measured grain boundaries was 4±2µm in PCM and the distance between the maxima and minima of eight grain boundaries in DICM amounted to 6±4µm. Also in DICM occasional pockets of fluid were identifiable. At t=2000min a narrow distribution of average crystallite size was observed with most crystals measuring between 60µm and 90µm and the maximum observed size being about 110µm. Analysis of the Bragg micrographs with their improved statistics corroborate this finding. The average value $<L>_M$ = (76±9)µm is only slightly larger than the average crystallite size obtained at that time from Bragg scattering. The observation of melt pockets at the intersection of the grain boundaries implies an important additional information: the sample definitely was prepared at the upper end of the coexistence range. We performed several scans of the sample to find that together with the grain boundaries (taken with an average width of 5µm) they make up about 5% of the sample volume and therefore the sample shows an approximate crystallinity X = 0.95 ± 0.05.

**Discussion**

To summarize the evaluation results of the last section, we observe for the Bragg regime: a) nucleation of crystals at a roughly constant rate for most of the conversion stage (c.f. Fig. 5b); b) an increase of crystallite dimensions with a power law exponent of 0.28 for all times $t<10^5\tau$, later decreasing to 0.22 (c.f. Fig. 4b); c) a saturation of the fraction of crystals after some $10^3\tau$ at a vale close to 95% (cf. Fig. 4b); d) a dominance of ripening for times $t>700\tau$ characterised by a power law decrease of the crystallite number density (c.f. Fig. 5a; and e) a constant increase of the crystal lattice constant (c.f. Fig. 4a), which is more rapid during the formation of the percolated crystal network. In addition we find from SALS: f) a SALS peak emerging only after $2\,10^4\tau$ (c.f. Fig. 6a and b) with a maximum intensity growing roughly linearly in time (c.f. Fig. 4c); g) an increase of the characteristic dimension of the SALS objects strictly parallel to that of the crystals (c.f. Figs 4c and d); and h) Furukawa scaling of the SALS peak (c.f. Fig. 7) with a fractal dimension of the scattering objects decreasing from 2.25 to 1.7 (c.f. Fig. 8). From microscopy we could show that j) the grain boundaries have the larger index of refraction but k) show less Bragg scattering than the crystalline regions. We further observed l) that the crystallite sizes from microscopy and Bragg scattering are smaller than the characteristic length of the SALS objects. And finally, l), we could identify melt pockets at the intersection of grain boundaries (c.f. Fig. 10), showing that the sample was prepared at the upper end of the coexistence range with a crystallinity of approximately 0.95. We now will discuss the origin of the SALS signal, compare our findings on the crystallization kinetics to other colloidal systems, and finally point out an interesting morphological analogy between our system and coarsening foams.

**Origin and interpretation of the SALS peak**

The late emergence of the SALS peak from the underlying, monotonically decaying signal is an unexpected result of our scattering experiments. For SALS the typical length scale by far exceeds that of the typical interparticle spacing. Possible origins for a SALS peak in colloidal systems are e.g. crystals, grain boundaries or depletion zones. At the investigated $q$ range the internal structure of these objects becomes irrelevant. Rather the forward bound scattering is either induced by a difference of the refractive index of the scatterers as compared to the average one or it is caused by objects attenuating the transmitted light. According to Clausius-Mosotti, the scattering power for a scatterer with refractive index $\nu$ immersed in a surrounding of average refractive index $\bar{\nu}$ is proportional to $(\nu^2 - \bar{\nu}^2)/(\nu^2 + 2\bar{\nu}^2) \cong (\nu - \bar{\nu})$, where the approximation is valid for conditions close to index match. Since the refractive index of a given colloidal phase is related to the volume fraction of particles in that phase, a SALS contrast may originate from fluctuations of particle volume fraction.

A difference in particle concentration in between different regions of the sample will also affect the transmission, as each particle scatters also in the high-$q$ Bragg range which is not detected in SALS. For disordered systems one would expect a decrease of transmission with increased concentration. For ordering systems the transmission may further depend on the details of the scattering patterns of both phases in the large $q$ regime and is much more difficult to estimate. To identify our scatterers we turn to the micrographs in Figs. 9 and 10. These are constructed from the superimposed light of zeroth and higher order diffracted beams. The angular range of scattered light collected for image construction coincides with that of our SALS experiment, except that in the SALS experiment there is a lower cut off at very small angles. Images taken on melt systems or at short times do not show any significant deviations from plain grey. In the images taken at large times crystallites, grain boundaries and melt pockets are identified. The combined analysis by PCM and DICM revealed that these are both phase and amplitude objects. The grain boundaries have the larger index of refraction but are less attenuated due to Bragg scattering than the crystalline regions. Both is compatible with the observed changes in volume fraction, where we find compacted crystals (c.f. Fig. 4a) which necessarily have to coexist with an expanded melt. As after conversion the crystalline phase occupies the larger fraction of the sample, its refractive index $\nu_X$ is closer to the average one $\bar{\nu}$. The larger deviations are present for the grain boundaries and melt pockets. Consequently it is the shape and the structural arrangement of the minority material that determines the shape of the SALS peak observed in our colloidal solid.

This conclusion allows to understand a few other observations. First, in Fig. 4d the characteristic SALS length scale was inferred to be on the order of 50-150µm, which was slightly larger than the crystallite size inferred from Bragg scattering and microscopy. This is attributed to a geometric effect. The grain boundaries are located outside the crystals and hence their distance is larger than the crystal diameter.

Second, the fitted dimensionality of the scattering objects changed from 2.25 at early times to 1.65 at late times. Microscopy, however, had shown that at all stages relevant for SALS the directly observable objects (crystallites and grain boundaries and melt pockets) are compact and not fractal in the classical sense. Therefore the values of $d_f$ have to be interpreted as average dimensionalities of grain boundaries and melt pockets. We therefore think that the observed decrease of dimensionality is connected to and may be explained with the evolution of the grain boundary shapes with time. The micrographs at t = 2000min showed the existence of (triangular) pouches of remaining melt, which occur at the intersection of plate-like grain boundaries. In the course of time the originally thick plate-like boundaries between the crystallites straightened and thinned. Further, the average crystallite size and hence the grain boundary spacing and the grain boundary intersection spacing increased by coarsening processes, which are structurally self similar as seen from the Furukawa scaling. However, while the network structure was retained, the melt was drained from the plate-like boundary regions into the intersection regions between these plates. Thereby the dominant shape of the grain boundaries changes from thick plates to thin plates plus rod-like intersections.

**Comparison to other systems**

We first compare our kinetic observations to those made on other colloidal systems. First, for times larger than one minute, nucleation proceeded at roughly constant nucleation rate density. This is in line with earlier observations on pure hard sphere systems [12, 13, 14, 15]. Unfortunately the present experiments have no sufficient temporal resolution to perform a comparison to more recent studies on hard sphere systems with different polydispersities. There a two step nucleation process was observed with an early precipitation of dense (amorphous?) objects which partially convert to crystals at later time and in parallel to secondary nucleation of crystals [45, 48]. Also for charged spheres slightly above melting a constant nucleation rate density was observed. At larger particle concentrations, however, the rate density appeared to become peaked during the early stage of conversion. The authors attributed this effect to the low compressibility of the melt, which dilutes as the compacted crystallites are formed and grow. Hence the peaked nucleation rate densities are related to a decreased super-saturation of the remaining melt.

Second, for our hard sphere polymer mixture growth was observed to follow a power law $\langle L \rangle \propto \tau^\alpha$ with an exponent of $\alpha_L = 0.28 \pm 0.015 \approx 1/3$, while in previous studies on hard spheres conversion exponents between 0.5 and 1 were observed for hard spheres [5, 12, 14, 15, 47]. These were interpreted in terms of diffusion limited and reaction controlled growth in the limits of $\tau^{1/2}$ and $\tau^1$. Intermediate exponents were attributed to long lived transient stages of pseudostatic growth [17]. A $\tau^{1/3}$ growth law corresponding to classical Ostwald ripening has been observed in HS samples prepared close to melting only during coarsening [12, 15] but not during conversion. In principle, both exponents may be lowered with increased $\Phi$ by polydispersity induced fractionation [49, 50, 51, 52]. In fact, recent systematic measurements on fractionating HS Microgels of polydispersity 0.068 showed a crossover of the conversion and coarsening exponents at $\Phi = 0.557 - 0.565$ and values of $\alpha_L \cong 0.1$ [30]. At coexistence, however, also these systems showed conversion exponents around $\alpha_L \approx 0.5$. We therefore conclude that in our case the observed exponent should not be influenced by polydispersity.

Charged spheres show reaction controlled growth with $\alpha_L = 1$ above coexistence [6], while across coexistence growth does not follow a power law. Rather, the initially linear growth slows and decays to zero as the equilibrium crystallinity X is reached [53]. This was attributed to a decreasing super-saturation of the remaining melt as it relaxed towards the equilibrium fluid at coexistence.

Thus the power law observed here during the conversion stage of a hard sphere polymer mixture differs from previously observed cases in the exponent value and in its origin. Furher, the robustness of the growth exponent of about 1/3 for $\langle L(t) \rangle$ over both conversion and coarsening stage are theoretically expected only for a conserved order parameter [36, 37, 38, 39]. In our recent Letter we suggested to explain this behaviour by explicitly considering our sample as a two component system composed of particles and polymer. When two particles approach to attain their equilibrium position, the polymer component has to leave the region between their surfaces. Any further particle which approaches the formed pair will encounter an increased polymer concentration in the vicinity of the pair. The process repeats and still larger polymer concentrations are generated around the growing clusters. Given a two phase state at equilibrium with $\Phi_X > \Phi_M$, balance of osmotic pressure requires diffusive transport of polymer away from the clusters into the phase depleted of particles. As a strictly local process, the successive addition of particles is independent of the amount of converted material. Therefore it also applies for the coarsening regime, where particles detach from the facet of a given crystal and independently others attach at the opposing surface of a second one. In this process the order parameter crystallinity is not conserved, but the total polymer concentration is. We therefore concluded that in colloid polymer mixtures with large size difference, the total polymer number density may become the controlling conserved order parameter and lead to a power law growth behaviour with an exponent of 1/3 [35].

From the point of theory, such a scenario is *not* expected, if the dispersion is described as an effective one component system. Still, this simplification is usually employed in calculations of the equilibrium properties of hard sphere plus polymer mixtures as well as in investigations addressing their crystallization behaviour [6, 30]. The coupling of the crystallization to the polymer density redistribution shows the importance of many body interactions and the necessity to explicitly consider all species in the theoretical or simulational modelling of phase transitions.

Ripening in our case started already before conversion was completed. Also in another study under µ-gravity the presence of ripening already during the conversion stage, i.e. while $X < 1$, had been observed in a sample prepared at coexistence [17]. At present it is not clear, whether the observation of ripening during the conversion stage bears a systematic correlation with the absence of crystallite settling, which may tend to alter crystallite sizes and morphologies. Ackerson an co-workers reported that settling shears away outer crystal layers in particular for large crystallites [18]. Therefore settling may possibly compensate the growth of large crystals upon the expense of small ones. In our case we had prepared the suspension very close to density match and no significant settling was observable during the experiments.

As mentioned, our coarsening exponent of 1/3 is in good agreement with the expectation for classical Ostwald ripening, with either the polymer density or the crystallinity taken as conserved order parameters [37, 38, 39]. An interesting point is the observed transient decrease of the exponent at rather late times. It is compatible with the expectation that elastic stress accumulation hinders further coarsening [54]. Coarsening exponents below 1/3 have also been observed in previous studies on HS [12, 13, 14, 15, 16, 17, 18, 19]. As discussed above, a decreased coarsening exponent may be partly due to polydispersity. But it also would be consistent with crystal-crystal straining occurring upon contact during conversion or build up in rearrangements during coarsening. In particular for samples prepared above melting, where the crystallinity is one, all crystallites are in intimate contact. The contact of differently oriented crystal faces then yields the friction necessary for the build up of stress. In our case at the upper end of the coexistence range, the crystallites appear (partially) lubricated by the melt filled grain boundaries. We therefore possibly profit from the larger "mobility" of crystallites within the remaining melt and the option to expel odd particles into the surrounding disordered phase. We believe that a lubrication effect may explain the long coarsening time before the build up of stress is reflected in the coarsening exponent and also the shortness of the low exponent interval. No coarsening data are as yet available for charged spheres.

**Analogy to foam systems**

Our experimental SALS signal appears after the formation of grain boundaries and their arrangement on a typical length scale as the crystallites get more monodisperse. Since their thickness is small as compared to the enclosed crystal, the characteristic length scale of the grain boundary network is very close to the crystallite size and similar kinetics are observed in the Bragg and SALS experiments. To the best of our knowledge, this is the first interpretation of SALS observations in terms of grain boundary scattering during melt crystallization or annealing of solids. We note, however, that also for pure HS this effect should be present and possibly has been seen already during the coarsening stage. He et al for instance observed a change of the scaling behaviour of the SALS data upon crystallite contact. It was discussed in terms of overlapping and finally vanishing depletion zones and also connected to crystallite break up [16]. It would therefore be interesting to combine SALS experiments on pure HS with microscopy and have a closer look at the origin of their late stage SALS signal.

An interesting consequence of our interpretation of the SALS signal is the remarkably close resemblance of our system to another class of soft matter systems, namely coarsening soap froths and other kinds of foams [55]. There as well the boundaries between air bubbles are in fact composed of plates and more rod-like objects at their intersection, the so-called Plateau-borders. Their thickness is controlled by the water content of a foam. For wet foams under gravity, the coarsening of air bubbles is unavoidably accompanied by a drying process in which first the lamella drain into the Plateau-borders, which in turn drain to the foam bottom until ultimately a dry foam is obtained. Interestingly, also the Plateau-border network structure is shape-persistent upon coarsening, meaning it scales in a self similar way with increased bubble sizes. In addition an increased spatial order may evolve during coarsening, as intersection angles close to 120° minimize surface energies.

The theoretical ground state of foams has been discussed by many authors, but the issue is not yet settled. Very early, Lord Kelvin (William Thompson) suggested a crystalline network of truncated octahedra (regular tetrakaidecahedra) with slightly bent facets to meet Plateau´s rules of intersection [56]. His conjecture, that this structure possesses the minimum surface area was disproved only after a century. Weaire and Phelan presented a counter-example composed of two different polyeder (an irregular tetrakaidecahedron and an irregular dodecahedron), which had a slightly lower surface area [57]. Also this structure was crystalline and showed slightly bent facets. Experimentally, body centred cubic packings of Kelvin cells have frequently been observed close to container walls or in restricted geometry, while the Weaire-Phelan structure was observed at the centre of a cylindrical sample [58, 59]. Both these structures are observed in well dried froths, while for wet froths a face centred cubic arrangement of spherical bubbles can be observed [60]. For wet froths a coarsening exponent of $\alpha_L = 0.32$ was reported, while for dry froths larger exponents were found (see the compilation in [55].) The theoretical equilibrated foam structures show crystalline symmetry. Experimentally, ordered foams have been hard to observe in bulk samples. The main reasons are the polydisperse bubble sizes, the lacking influences of templating container walls and the presence of gravity [60].

In our HS polymer mixture, the observed structure of the grain boundary network shows rather similar features, except that it is not "drying" as the mass densities of both the bubbles (crystallites), lamellae (grain boundaries) and Plateau borders (rod-like intersection regions) are very similar. Also our system is isotropic and does not display a long nor a short range order. (A Porod-plot of the scattered SALS intensity versus $q^4$ shows only one single peak.) Rather, we have a single length scale only, which corresponds to the average distance between opposing facets or Plateau borders. In this sense we have observed a wet, non-drying, polydisperse and disordered foam, which evolves towards monodispersity by diffusive displacement of the boundary material, but does not order. Possibly, this state is separated from a crystalline ground state by a considerable energy barrier.

**Conclusions**

We have investigated the solidification behaviour of a colloid polymer mixture at coexistence conditions by various time resolved optical methods. The present study significantly drew from the incidental choice of the nominal volume fraction at the upper end of the coexistence regime.

Above coexistence there would have been no second phase present as the sample solidifies completely. At lower $\Phi$ it would have been more difficult to identify the scattering objects from microscopic experiments. And these were the key to explain the SALS peak. Further, the experiments were carried out with the uncleaned samples containing the polymer from synthesis. Also this was not intended, but led to the significant deviations from the originally expected hard sphere behaviour. At present we follow the charming task to repeat the experiments with a Bragg scattering machine of still somewhat better temporal and angular resolution and with cleaned samples mixed with well defined polymers and also across the complete coexistence region.

Still, the present study has revealed a number of surprising results. We observed a late stage small angle signal of origin different to those formerly known from depletion zone scattering. We have carefully characterized the origin of this signal to find that it stems from the difference in grain boundary refractive index and average refractive index. With somewhat larger experimental sensitivity it should also be confirmable in other colloidal samples at the upper end of the coexistence region. Grain boundary scattering therefore may open another access to systematic investigations on coarsening kinetics and mechanisms.

As already stated in our Letter [35], we have conducted a study on a system in which two qualitatively different transitions compete, the first order freezing transition and a phase separation process. In particular, we have described a coupling of crystallization to the transport of a second polymeric component, which does not partake in crystallization. One main result was that the conserved order parameter polymer density controlled the kinetics of the combined process, rather than other order parameters like crystallinity (non conserved) or particle density (conserved) involved in the freezing transition and governing the kinetics in the pure HS case. The observed effect may be quite general and also have consequences for the formation kinetics of nematic or smectic phases from a hard plate-polymer mixture.

Finally, for our system at the upper end of the coexistence region we believe to have identified a three dimensional foam system hardly influenced by gravity. Our froth does not dry, but evolves at constant liquid content tuneable through the choice of particle volume fraction. This could provide an interesting alternative to µ-gravity experiments for the investigation of foam structure evolution.

**Acknowledgements**

It is a pleasure to thank K. Binder, W. Paul, T. Schilling and S. Iacopini for the many intense and fruitful discussions and J. Othegraven for the particle synthesis. Financial support of the DFG (Pa459/8; Pa459/12-14, and Pa459/16, SFB 428 and TR6 (print charge)), the EU MCRT-Network CT-2003-504 712, and the former MWFZ, Mainz is gratefully acknowledged.

**Appendix : Microscopic techniques**

For Phase Contrast Microscopy (PCM) an annular front aperture is used in combination with Köhler illumination, such that the sample is illuminated with a cone of parallel white light. An amplitude object reduces the transmitted intensity ($0^{th}$ order) while the $1^{st}$ order diffracted light is phase shifted by $\pi$. The amplitude of the diffracted wave is proportional to the refractive index difference of the object with respect to the average refractive index. A phase object leaves the transmitted amplitude unchanged, while the phase of the $1^{st}$ order diffracted light is retarded by $\pi/2$. In the back focal plane of the objective a phase ring is mounted, on which the $0^{th}$ order light is focused to create an image of the entrance annular aperture. The phase ring forward-shifts the phase of the illuminating beam by $\pi/2$ and furthermore attenuates the intensity. Light diffracted by the object passes the back focal plane without additional shift and attenuation. In the image plane the image is constructed from the superposition of $0^{th}$ order transmitted light (phase shift $+\pi/2$) and $1^{st}$ order diffracted light (phase shift $-\pi/2$) such that destructive interference results (negative phase contrast). Thus the phase information is transferred to an amplitude information. Thicker samples or larger refractive index differences result in large amplitudes for the diffracted light and thus in a larger image intensity than for thin objects of low refractive index difference. For amplitude objects the recombined $0^{th}$ and $1^{st}$ order light have a phase difference of $\pi/2$. Instead of interference a phase modulated wave results. The image appears darker, but otherwise identical to a conventional bright field image. Regions of low extinction appear brighter than regions of large extinction.

A different approach is followed in Differential Interference Contrast Microscopy (DICM), which is sensitive to lateral changes (gradients) of attenuation and refractive index. Here the object is illuminated by parallel white light (Köhler illumination). On the illumination side a polarizer generates linearly polarized light, which is split by a Wollaston prism into two parallel beams which are polarized normal to each other. The initial polarization is chosen with an angle of 45° with respect to the polarisation of the two resulting beams, such that equal intensities result. In addition, one of the beams is laterally shifted by some 30nm, which is well below the resolution of the objective. Past the objective a second Wollaston prism recombines the two beams (Nomarski configuration) which have sampled laterally separated regions of the object. An analyser is placed between prism and image plane which is oriented slightly off 90° with respect to the polarizer. If neither phase nor amplitude objects are present a grey image results (biasing). Objects of constant refractive index give the same phase shift to both laterally displaced beams. Thus no effect is seen after recombination. Objects of constant amplitude appear darkened proportional to the attenuation of the beams. Phase gradients retard the two beams passing the object differently and after recombination elliptically polarized light results. The direction of the gradient determines the rotation direction of the elliptically polarized light, but this effect is not resolved by the analyzer. In both cases an increased intensity results which for small differences in refractive index is proportional to the steepness of the gradient. In the case of differential attenuation of the two laterally shifted beams (amplitude object or scattering object) again linearly polarized light results with a turn of the orientation depending on the relative intensities of the beams. As the analyzer is oriented off 90°, this leads - as compared to the bias intensity - to an increased or decreased intensity which depends on the gradient of attenuation. In general both effects will appear simultaneously. By analyzing the symmetry of the intensity distribution across the object, DICM therefore may discriminate between the relative contributions of phase and amplitude effects.